\pdfoutput=1

\documentclass[aps,twocolumn,amsmath,amssymb,floatfix,longbibliography,prl]{revtex4-2} %,superscriptaddress

\usepackage[utf8]{inputenc}
\usepackage{newtxtext}
\usepackage[upint]{newtxmath}
\usepackage{microtype}
\usepackage{textcomp}
\usepackage{eucal}
\usepackage{bm}
\usepackage{siunitx}
\usepackage{comment}

\usepackage{accents}
\usepackage{mathtools}

\usepackage{enumerate}
\usepackage{amsfonts}
\usepackage{amsmath}
\usepackage{amssymb}
\usepackage{color}
\usepackage{soul}

\usepackage{graphicx}

\usepackage[colorlinks,allcolors=blue]{hyperref}
\usepackage[capitalize]{cleveref}

\definecolor{Red}{rgb}{1,0,0}

\newcommand{\ve}{\boldsymbol} % Vector

\newcommand{\prlsection}[1]{\textit{#1}.\kern0.05em---\kern0.05em\ignorespaces}

\begin{document}
\title{Nonreciprocal Josephson current through a conical magnet}

\author{Lina Johnsen Kamra}
\email{ljkamra@mit.edu}
\affiliation{Department of Physics, Massachusetts Institute of Technology, Cambridge, MA 02139, USA}
\author{Liang Fu}
\affiliation{Department of Physics, Massachusetts Institute of Technology, Cambridge, MA 02139, USA}

\date{\today}

%----------------------------------------------------

% Abstract:

\begin{abstract}

Superconductors can form ideal diodes carrying nondissipative supercurrents in the forward direction and dissipative currents in the backward direction. The Josephson diode has proven to be a promising design where the junction between the two superconductors comprises the weakest link and thus provides the dominant mechanism. We here propose a Josephson diode based on a single magnetic material with a conical spin structure. The helical spin rotation produces Rashba-like band splitting inversely proportional to the rotation period. Together with the Zeeman splitting caused by the time-reversal symmetry breaking of the noncoplanar spin texture, this results in a large diode efficiency close to the $0-\pi$ transition of the magnetic Josephson junction.

\end{abstract}

% Make title head:

\maketitle

\prlsection{Introduction}
The p-n junction is a prototypical example of a diode used in modern electronics, as it offers a small resistance in the forward direction and a much larger resistance in the backward direction. A breakthrough has been achieved recently with the realization of the superconducting ideal diode that admits zero-resistance nondissipative supercurrent flow in the forward direction and a finite-resistance dissipative current in the backward direction \cite{Ando_Nature_2020,Nadeem_NatRevPhys_2023,Nagaosa_AnnuRevCondensMatterPhys_2024}. To obtain such behavior, one exploits the nonreciprocity in the critical current $J_{\text{c}}$ -- the current bias at which the superconductor transitions between a nondissipative and resistive state \cite{Levitov_JETPLett_1985}. If the applied current $J$ is smaller than the critical current in the forward direction ($J<J_{\text{c},+}$) and larger than the critical current in the backward direction ($J>J_{\text{c},-}$), the system shows superconducting diode behavior. For the effect to be robust over a larger range of current biases, it is desirable to have a large diode efficiency $\eta=(J_{\text{c},+}-J_{\text{c},-})/(J_{\text{c},+}+J_{\text{c},-})$ \cite{Nagaosa_AnnuRevCondensMatterPhys_2024}.

Superconducting diode behavior can appear when the inversion symmetry and time reversal symmetry are broken \cite{Yuan_PNAS_2021,Daido_PRL_2022,He_NewJPhys_2022,Ilic_PRL_2022,Zinkl_PhysRevRes_2022}. There has been a flurry of activity demonstrating new mechanisms \cite{Misaki_PRB_2021,Davydova_SciAdv_2022,Zhang_PRX_2022,Suoto_PRL_2022,Legg_PRB_2022,Hu_PRL_2023,Sundaresh_NatCommun_2023,Hess_PRB_2023,Steiner_PRL_2023,dePicoli_PRB_2023,Gaggioli_arXiv_2024} for accomplishing this phenomenon \cite{Wu_Nature_2022,Baumgartner_NatNanotech_2022,Pal_NatPhys_2022,Jeon_NatMater_2022,Bauriedl_NatCommun_2022,Lin_NatPhys_2022,Narita_NatNanotech_2022,Suri_ApplPhysLett_2022,Hou_PRL_2023,Gutfreund_NatCommun_2023,Yun_PhysRevRes_2023,Trahms_Nature_2023,Ghosh_NatMater_2024} within only a couple of years. Within these efforts, the Josephson diode \cite{Wu_Nature_2022,Baumgartner_NatNanotech_2022,Pal_NatPhys_2022,Jeon_NatMater_2022,Misaki_PRB_2021,Davydova_SciAdv_2022,Zhang_PRX_2022} has proven to be a promising design where the dominating mechanism can more easily be distinguished as the Josephson junction itself comprises the weakest link in the system. Recent experiments on Josephson diodes in inversion symmetry breaking Rashba systems \cite{Pal_NatPhys_2022,Jeon_NatMater_2022}, have relied on time reversal symmetry breaking via an external magnetic field \cite{Pal_NatPhys_2022,Baumgartner_NatNanotech_2022} or proximity-coupling to a magnet \cite{Jeon_NatMater_2022}. 

In this work, we propose a Josephson diode that utilizes a single magnetic material, thus eliminating the need for both strong Rashba spin-orbit coupling and an external magnetic field.
In our proposed design, the magnetic layer between the two superconductors has a helical rotation of the spin-splitting field that gives rise to quasi-one-dimensional (quasi-1D) Rashba-like band splitting which breaks inversion symmetry. This splitting is inversely proportional to the period of the spin helix \cite{Meng_PRB_2019,Hals_PRB_2017}, and results in two well-separated Fermi surfaces. In addition, time-reversal symmetry breaking arises when tilting the helical spin-splitting field towards a conical spin structure. 
We analytically demonstrate how the asymmetric dispersion of the conical magnet gives rise to nonreciprocal critical currents. By numerically solving the Bogoliubov--de Gennes (BdG) equations, we find a large diode efficiency in the vicinity of the $0-\pi$ transition of the Josephson junction, achievable in conical magnets such as Ho \cite{Sosnin_PRL_2006,Robinson_Science_2010,Witt_PRB_2012}. Furthermore, a magnetic field can be applied to induce and control spin-canting in helimagnets such as Cr$_{1/3}$NbS$_2$ \cite{Cao_MaterTodayAdv_2020,Yonemura_PRB_2017,Spuri_PhysRevResearch_2024} and thereby tune the diode effect. 

\begin{figure}[t]
    \centering
    \includegraphics[width=\columnwidth]{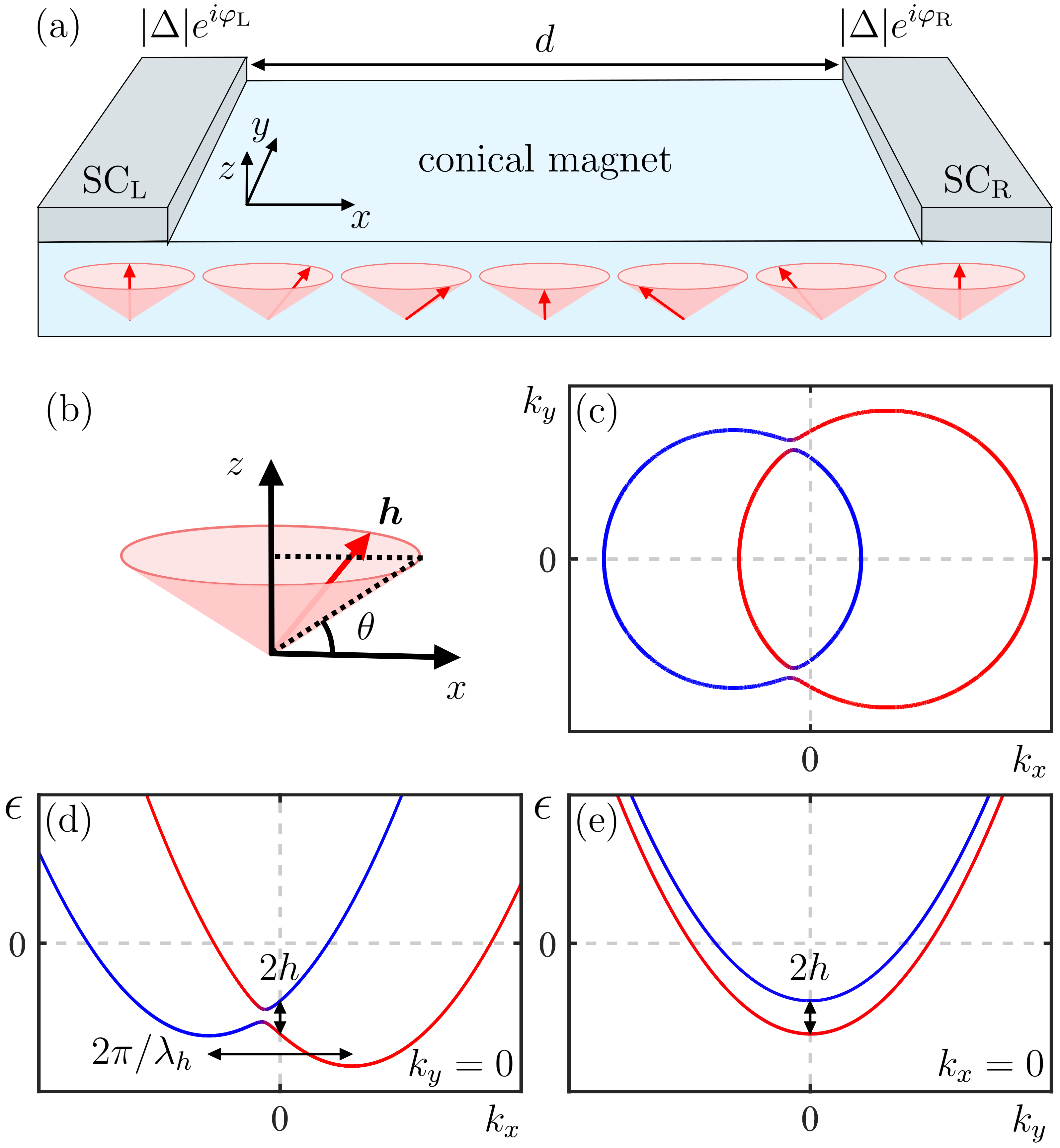}
    \caption{(a) A Josephson junction between two superconductors, SC$_{\text{L}}$ and SC$_{\text{R}}$, with the same gap $|\Delta|$ has a phase difference $\Delta\varphi=\varphi_{\text{R}}-\varphi_{\text{L}}$. SC$_{\text{L}}$ and SC$_{\text{R}}$ are separated by a distance $d$ and connected via a conical magnet. (b) The local spin-splitting field of the conical magnet is of magnitude $h$, has a rotation period $\lambda_h$, and its canting is described by the angle $\theta$. (c) The Fermi surface and (d)-(e)~the corresponding normal-state dispersion [Eq.~\eqref{eq:eienenergies}] of the conical magnet (exaggerated for clarity) is nonreciprocal under inversion of $k_x$ [panel~(d)] and symmetric under inversion of $k_y$ [panel~(e)]. Blue (red) curves correspond to spin up (down). The helical spin rotation causes a horizontal shift in the dispersion [panel~(d)] through a quasi-1D Rashba-like inversion symmetry breaking [Eq.~\eqref{eq:H2}], as well as the avoided band crossing [panel~(d)] separating the Fermi surface into two distinct lobes [panel~(c)]. The vertical Zeeman band splitting results from the canting.}
    \label{fig:01}
\end{figure}

\prlsection{Nonreciprocal dispersion}
We consider a Josephson junction where a supercurrent runs between two superconductors (SC$_{\text{R}}$ and SC$_{\text{L}}$) due to a phase difference $\Delta\varphi=\varphi_{\text{R}}-\varphi_{\text{L}}$. As the supercurrent runs through the metallic magnet connecting the two superconductors, it is subjected to a conical spin-splitting field, see Fig.~\ref{fig:01}(a)-(b). To demonstrate how this spin-splitting field can give rise to nonreciprocity, we first consider the normal-state band structure of the conical magnet.

The conical magnet can be described by a Hamiltonian \cite{Meng_PRB_2019}
\begin{align}
    H_{\text{cone}} = &\int d\ve{r}\:\sum_{\sigma}\psi_{\sigma}^{\dagger}(\ve{r})\left(-\frac{\hbar^2\nabla_{\ve{r}}^2}{2m}-\mu\right)\psi_{\sigma}(\ve{r})\notag\\
    +&\int d\ve{r}\:\sum_{\alpha,\beta}\psi_{\alpha}^{\dagger}(\ve{r})\left[\ve{h}(x)\cdot\ve{\sigma}\right]_{\alpha,\beta}\psi_{\beta}(\ve{r}), \label{eq:H1}
\end{align}
where $\psi_{\sigma}^{(\dagger)}(\ve{r})$ annihilates (creates) an electron of spin $\sigma$ at position $\ve{r}=(x,y)$, $m$ is the electron mass, $\mu$ is the chemical potential, and $\ve{\sigma}$ is the vector of Pauli matrices. The spin space and real space coordinates of the above Hamiltonian are completely decoupled. Without loss of generality, we consider a local spin-splitting field
\begin{align}
    \ve{h}(x) &=h\cos(\theta)\{\hat{x}\sin\left[\phi(x)\right]+\hat{y}\cos\left[\phi(x)\right]\}\notag\\
    &+h\sin(\theta)\hat{z}.
    \label{eq:h}
\end{align}
tilted by an angle $\theta$ towards the $z$ axis [Fig.~\ref{fig:01}(b)]. Its rotation around the same axis is described by the angle $\phi(x)$. We assume a monotonous spin rotation so that $\partial_x\phi(x)=2\pi/\lambda_h$ is constant. The parameter $\lambda_{h}$ is the length scale of a $2\pi$ rotation of the spin-splitting field. 
To eliminate the position dependence of the spin-splitting field [Eq.~\eqref{eq:h}] from the Hamiltonian [Eq.~\eqref{eq:H1}], we perform a unitary transformation $U(x)=\text{exp}[-i\phi(x)\sigma_z/2]$ to a rotating reference frame \cite{Hals_PRB_2017,Hess_PRB_2022}. The resulting Hamiltonian,
\begin{align}
    H_{\text{cone}} = &\int d\ve{r}\:\sum_{\sigma}\tilde{\psi}^{\dagger}_{\sigma}(\ve{r})\left(-\frac{\hbar^2\nabla_{\ve{r}}^2}{2m}-\tilde{\mu}\right)\tilde{\psi}_{\sigma}(\ve{r})\notag\\
    +&\int d\ve{r}\:\sum_{\alpha,\beta}\tilde{\psi}^{\dagger}_{\alpha}(\ve{r})\left(\tilde{\ve{h}}\cdot\ve{\sigma}-i\tilde{\alpha}\sigma_z\partial_x\right)_{\alpha,\beta}\tilde{\psi}_{\beta}(\ve{r}), \label{eq:H2}
\end{align}
where $\tilde{\psi}_{\sigma}(\ve{r})=U(x)\psi_{\sigma}(\ve{r})$, has a position independent spin-splitting field 
$\tilde{\ve{h}}=h[\hat{y}\cos(\theta)+\hat{z}\sin(\theta)]$, 
where the helical spin rotation has been mapped onto a constant spin-splitting field along $\hat{y}$ and a quasi-1D Rashba-like inversion symmetry breaking along $\hat{x}$ of magnitude $\tilde{\alpha}=(\hbar^2/2m)(2\pi/\lambda_{h})$. The transformation also shifts the chemical potential to $\tilde{\mu} = \mu-(\tilde{\alpha}/2)^2/(\hbar^2/2m)$. The Hamiltonian in Eq.~\ref{eq:H2} resembles that of a quasi-1D Rashba nanowire under an applied spin-splitting field -- a minimal model for achieving a superconducting diode effect \cite{Legg_PRB_2022,dePicoli_PRB_2023}. 

To study the effect of the quasi-1D Rashba-like inversion symmetry breaking and uniform spin-splitting field, we apply the Fourier transform $\tilde{\psi}_{\sigma}(\ve{r})=\int[d\ve{k}/(2\pi)^2]\:\tilde{\psi}_{\sigma}(\ve{k})\text{exp}(i\ve{k}\cdot\ve{r})$ to Eq.~\eqref{eq:H2} and diagonalize the Hamiltonian.
%The resulting Hamiltonian is given by 
%\begin{align}
%    H_{\text{cone}} = &\int \frac{d\ve{k}}{(2\pi)^2}\:\sum_{\sigma}\tilde{\psi}^{\dagger}_{\sigma}(\ve{k})\left(\frac{\hbar^2k^2}{2m}-\tilde{\mu}\right)\tilde{\psi}_{\sigma}(\ve{k})\notag\\
%    +&\int \frac{d\ve{k}}{(2\pi)^2}\:\sum_{\alpha,\beta}\tilde{\psi}^{\dagger}_{\alpha}(\ve{k})\left(\tilde{\ve{h}}\cdot\ve{\sigma}+\tilde{\alpha}k_x\sigma_z\right)_{\alpha,\beta}\tilde{\psi}_{\beta}(\ve{k}). \label{eq:H2b}    
%\end{align}
The resulting dispersion \cite{Meng_PRB_2019},
\begin{align}
    \epsilon_{\pm}(\ve{k})=&\left(\frac{\hbar^2k^2}{2m}-\tilde{\mu}\right)\notag\\
    &\pm\tilde{\alpha}\sqrt{\left[k_x+\frac{h\sin(\theta)}{\tilde{\alpha}}\right]^2+\left(\frac{h\cos(\theta)}{\tilde{\alpha}}\right)^2},
    \label{eq:eienenergies}
\end{align}
and Fermi surface, at which the dispersion of the conical magnet crosses the Fermi energy [$\epsilon_{\pm}(\ve{k}_{\text{F}})=0$], is plotted in Fig.~\ref{fig:01}(c)-(e). The quasi-1D Rashba-like inversion symmetry breaking shifts the spin-up (spin-down) energy band towards smaller (larger) $k_x$ with a relative shift between the two bands of $2\pi/\lambda_h$. Importantly, the shift is inversely proportional to $\lambda_h$ enabling the design of a large quasi-1D Rashba-like band splitting in conical magnets with a short rotation period, typically of the order of nanometers or tens of nanometers in Ho \cite{Witt_PRB_2012} and Cr$_{1/3}$NbS$_2$ \cite{Togawa_PRL_2012}, respectively. Note that the system is invariant under an equal rotation of all spins, so that a $\phi$ varying along $\hat{n}$ results in a quasi-1D Rashba-like splitting with respect to $\ve{k}\cdot\hat{n}$. This decoupling of the spin space and real space enables the Josephson junction to have nonreciprocal critical currents along the along the net spin-splitting field \cite{Gaggioli_arXiv_2024}.

The $\hat{y}$ component of $\tilde{\ve{h}}$, associated with the helical spin rotation, gives rise to an avoided band crossing between the two bands that separates the Fermi surface into two distinct lobes. The $\hat{z}$ component, arising from the out-of-plane tilt, gives a relative vertical shift between the two bands. At $k_x=0$, or when approaching ferromagnetic alignment $(\theta\to\pi/2)$, the relative shift between the bands is of magnitude $2h$. The combination of inversion and time-reversal symmetry breaking from the quasi-1D Rashba-like and Zeeman band splitting, respectively, result in the Fermi surfaces and dispersion being asymmetric under inversion of $k_x$.

\begin{figure}[t]
    \centering
    \includegraphics[width=\columnwidth]{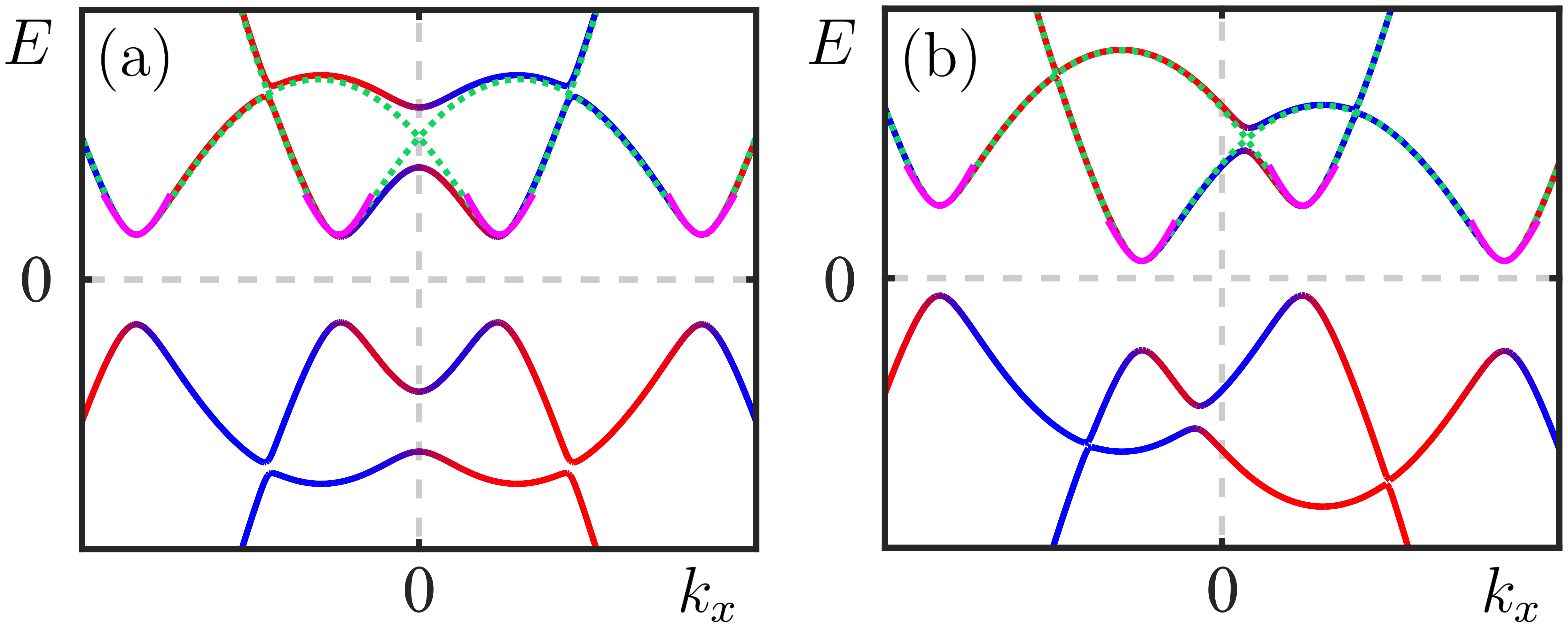}
    \caption{Proximity-induced superconductivity opens a gap of magnitude $2|\Delta|$ in the dispersion of the conical magnet [Fig.~\ref{fig:01}(d)]. (a) For a helical magnet ($\theta=0$), the gap is centered around $E=0$ and the dispersion is symmetric. (b) For a conical magnet (here $\theta=0.75\pi/2$), the gap is shifted by $+(-)h\sin(\theta)$ at $ k_x^{\text{F,i}}$ and $-k_x^{\text{F,o}}$ ($ k_x^{\text{F,o}}$ and $-k_x^{\text{F,i}}$). The analytic expressions in Eqs.~\eqref{eq:Eanalytic} and~\eqref{eq:Eanalytic2} correspond to the green and magenta curves, respectively.}
    \label{fig:01b}
\end{figure}

To understand the behavior of Cooper pairs formed from electrons within one Fermi pocket, we estimate the difference
\begin{align}
    \Delta k_x^{\text{F,o(i)}}=+(-)\frac{h\sin(\theta)\sqrt{2m/\hbar^2}}{\sqrt{\tilde{\mu}+(\hbar^2/2m)(\pi/\lambda_{h})^2}}
\end{align}
between the Fermi momenta in the positive and negative direction on the outer (inner) pocket [Fig.~\ref{fig:01}(c)]
for $k_y = 0$. We assumed that $h\ll\tilde{\mu}$ and neglected $h\cos(\theta)$ in Eq.~\eqref{eq:eienenergies}. The latter only affects the dispersion at the avoided band crossing away from the Fermi energy or at large $k_y$. The shift $\Delta k_x^{\text{F,o(i)}}$
has opposite signs on the outer and inner pocket, while the Fermi velocity $v_x^{\text{F,o(i)}}=\pm'(1/\hbar)\partial_{k_x}E_{\pm,\pm'}(k_x,k_y=0)\big|_{k_x = k_x^{\text{F,o(i)}}}$ take different values
\begin{align}
\left|v_x^{\text{F,o(i)}}\right|=\sqrt{\frac{2}{m}}\left[\sqrt{\tilde{\mu}+\frac{\hbar^2}{2m}\left(\frac{\pi}{\lambda_h}\right)^2}-\frac{1}{2}\sqrt{\frac{\hbar^2}{2m}}\Delta k_x^{\text{F,o(i)}}\right],
\end{align}
on the outer and inner pocket due to the opposite signs of $\Delta k_x^{\text{F,o(i)}}$.
When $h\sin(\theta)>0$, the coherence length $\xi_0^{\text{o(i)}}\sim v_x^{\text{F,o(i)}}$ of Cooper pairs formed from electrons in the outer (inner) pocket is therefore reduced (increased) by the out-of-plane tilt. Thus, while the contributions from the two Fermi pockets partly compensate each other owing to the opposite signs of $\Delta k_x^{\text{F,o(i)}}$, the contribution from the inner pocket dominates due to the increased coherence length. 

To estimate the resulting nonreciprocity in the superconducting gap $\Delta$, we include proximity-induced superconductivity via the Hamiltonian 
\begin{align}
    H_{\text{SC}}=-\int \frac{dk_x}{2\pi}\:\left[\Delta\:\tilde{\psi}_{\downarrow}^{\dagger}(k_x)\tilde{\psi}^{\dagger}_{\uparrow}(-k_x)+\text{h.c.}\right].
\end{align}
The positive eigenenergies of the total Hamiltonian $H=H_{\text{cone}}+H_{\text{SC}}$ are given by
\begin{align}
    E_{\pm}(k_x)=&\sqrt{\left\{\frac{\hbar^2}{2m}\left[k_x\pm\frac{\pi}{\lambda_h}\right]^2-\left[\tilde{\mu}+\frac{\hbar^2}{2m}\left(\frac{\pi}{\lambda_h}\right)^2\right]\right\}^2+\left|\Delta\right|^2}\notag\\
    &\pm h\sin(\theta),
    \label{eq:Eanalytic}
\end{align}
again assuming $h\ll\tilde{\mu}$ and neglecting $h\cos(\theta)$. We write the above eigenenergies in terms of $k_x = k_x^{\text{F,hel}}+\delta k_x$, where $\delta k_x$ is a small deviation away from the Fermi momentum $k_x^{\text{F,hel}}$ of a helix $(\theta=0)$ with magnitude $\big|k_x^{\text{F,o(i)}}\big|=\sqrt{2m/\hbar^2}\sqrt{\tilde{\mu}+(\hbar^2/2m)(\pi/\lambda_h)^2}+(-)\pi/\lambda_h$. To the lowest order in $\delta k_x$ \cite{Davydova_arXiv_2024}, the dispersion on the outer (inner) pocket is
\begin{align}
    E^{\text{o(i)}}(k_x)=\sqrt{\big(v_x^{\text{F,hel}}\delta k_x\big)^2+\left|\Delta\right|^2}-(+)\text{sgn}(k_x)h\sin(\theta),
    \label{eq:Eanalytic2}
\end{align}
where $v_{x}^{\text{F,hel}}=\sqrt{2\hbar^2/m}\sqrt{\tilde{\mu}+(\hbar^2/2m)(\pi/\lambda_h)^2}$ is the Fermi velocity of a helix. At $\delta k_x = 0$, the superconducting gap $\Delta$ is shifted by $\pm h\sin(\theta)$ as shown in Fig.~\ref{fig:01b}. The critical current of intra-pocket Cooper pairs is therefore nonreciprocal \cite{Yuan_PNAS_2021}.

\prlsection{Numerical method}
To calculate the resulting diode efficiency, we start by discretizing the Hamiltonian in Eq.~\eqref{eq:H1} onto a square lattice of size $N_x\times N_y$ in the $xy$ plane, 
\begin{align}
    H = &
    -t\sum_{\langle\ve{i},\ve{j}\rangle,\sigma}\left(c_{\ve{i},\sigma}^{\dagger}c_{\ve{j},\sigma}+\text{h.c.}\right)
    -\mu\sum_{\ve{i},\sigma}c_{\ve{i},\sigma}^{\dagger}c_{\ve{i},\sigma}\notag\\
    &-\sum_{\ve{i}}\left(\Delta_{i_x}c_{\ve{i},\downarrow}^{\dagger}c_{\ve{i},\uparrow}^{\dagger}+\text{h.c.}\right)\notag\\
    &+\sum_{\ve{i},\alpha,\beta}c_{\ve{i},\sigma}^{\dagger}\left(\ve{h}_{i_x}\cdot\ve{\sigma}\right)_{\alpha,\beta}c_{\ve{i},\beta}, \label{eq:H3}
\end{align}
where the continuum electron operators $\psi_{\sigma}(\ve{r})$ have been replaced with $c_{\ve{i},\sigma}$, where $\ve{i}=(i_x,i_y)$ is the lattice site index. Similarly, the spin-splitting field $h(x)$ [Eq.~\eqref{eq:h}] and the associated rotation angle $\phi(x)$ have been replaced by the discrete $h_{i_x}$ and $\phi_{i_x}$. The length scale of the spin rotation is assumed to take discrete values $\lambda_h/a$, where $a$ is the lattice constant. Electrons can hop between nearest neighbor sites with a hopping parameter~$t$. We have introduced conventional on-site superconducting pairing assuming that a superconducting gap $\Delta_{i_x}$ is proximity-induced onto the first and last $10$ sites along the $x$ axis only. At these sites, it takes the value \cite{Tinkham_book_1996}
\begin{align}
    \Delta(T) = \Delta_0\tanh\left(1.74\sqrt{\frac{T_\text{c}}{T}-1}\right)e^{i\varphi_{\text{L(R)}}},
\end{align}
where $T_{\text{c}} = \Delta_0/1.76$ is the superconducting critical temperature. We scale all length scales with respect to the coherence length of the superconductors
$\xi_0=\hbar v_{\text{F}}/\pi\Delta_0$, where
$v_{\text{F}}=(1/\hbar)|\nabla_{\boldsymbol{k}}E(\boldsymbol{k})|_{\boldsymbol{k}=\boldsymbol{k}_{\text{F}}}$
is the Fermi velocity in the normal-state, where the dispersion is given by $E(\boldsymbol{k})=-2t[\cos(k_x)+\cos(k_y)]-\mu$.

Assuming periodic boundary conditions along $\hat{y}$, we numerically diagonalize the Hamiltonian in Eq.~\eqref{eq:H3} by solving the BdG equations \cite{Zhu_Book_2016,Chourasia_PRB_2023} [see the Supplemental Material (SM) \footnote{See the Supplemental Material (SM) for an outline of the numerical method, and additional results for the average critical current at different values of the local spin-splitting field $h$ and the diode efficiency for different Fermi energies.}]. 
In order to approach the dispersion in Fig.~\ref{fig:01}(d)-(e), we consider a fixed filling fraction 
$f = (1/2N_xN_y)\sum_{\ve{i},\sigma}\langle c_{\ve{i},\sigma}^{\dagger}c_{\ve{i},\sigma}\rangle$
well below half-filling. The Fermi energy $E_{\text{F}}=2t+\mu$ is defined as the energy from the bottom of the normal-state bulk band at $h=0$.

To find the current-phase diagram of the magnetic Josephson junction, we calculate the average of the local bond current
    $J^x_{i_x+1,i_x} = it\sum_{\sigma}\langle c_{i_x+1,\sigma}^{\dagger}c_{i_x,\sigma}-\text{h.c.}\rangle$
from site $i_x$ to site $i_x+1$ inside the region where $\Delta(T)=0$ \cite{Zhu_Book_2016}. 
The data is fit to the current-phase relation \cite{Golubov_RevModPhys_2004}
\begin{align}
    J(\Delta\varphi)=J_1\sin(\Delta\varphi-\varphi_1)+J_2\sin[2(\Delta\varphi-\varphi_2)].\label{eq:Jphi}
\end{align}
For $\varphi_1 = 0$, the first term corresponds to the conventional Josephson current found in normal-metal junctions. An anomalous phase shift $\varphi_1$ allows for $0$-$\pi$ transitions when $\varphi_1$ transitions from $0$ to $\pi$ as observed in ferromagnetic Josephson junctions \cite{Buzdin_JETPLett_1991,Ryazanov_PRL_2001,Birge_APLMater_2024}. Intermediate phase shifts have been predicted in the presence of Rashba spin-orbit coupling \cite{Buzdin_PRL_2008,Bergeret_EurophysLett_2015}, for noncollinear ferromagnets \cite{Kulagina_PRB_2014,Silaev_PRB_2017} and conical spin structures 
\cite{Meng_PRB_2019}, but these works did not include the second harmonic term that givs rise to a diode effect. 

To quantify the superconducting diode effect, we first find the maximum critical current $|J_{\text{max}}(\Delta\varphi_{\text{max}})|=\text{max}|J(\Delta\varphi)|$ and for $\Delta\varphi_{\text{max}}>(<)0$ define the critical current for positive and negative $\Delta\varphi$ as 
\begin{align}
    J_{\text{c},+(-)}&=+(-)J(\Delta\varphi_{\text{max}}),\\
    J_{\text{c},-(+)} &= -(+)\text{min}\{\text{sgn}[J_{\text{c},+(-)}]J(\Delta\varphi)\Theta[-(+)\Delta\varphi]\}.
\end{align} 
This definition allows us to distinguish between $0$ and $\pi$ junctions where $J_{\text{c},+},J_{\text{c},-}>(<)0$ for the former (latter).
When $J_{\text{c},+}\neq J_{\text{c},-}$, the Josephson junction behaves as a superconducting diode with efficiency $\eta = (J_{\text{c},+}-J_{\text{c},-})/(J_{\text{c},+}+J_{\text{c},-})$ \cite{Nagaosa_AnnuRevCondensMatterPhys_2024}.
A finite diode efficiency requires the second harmonic term in Eq.~\eqref{eq:Jphi} to be finite.

\begin{figure}[tb]
    \centering
    \includegraphics[width=\columnwidth]{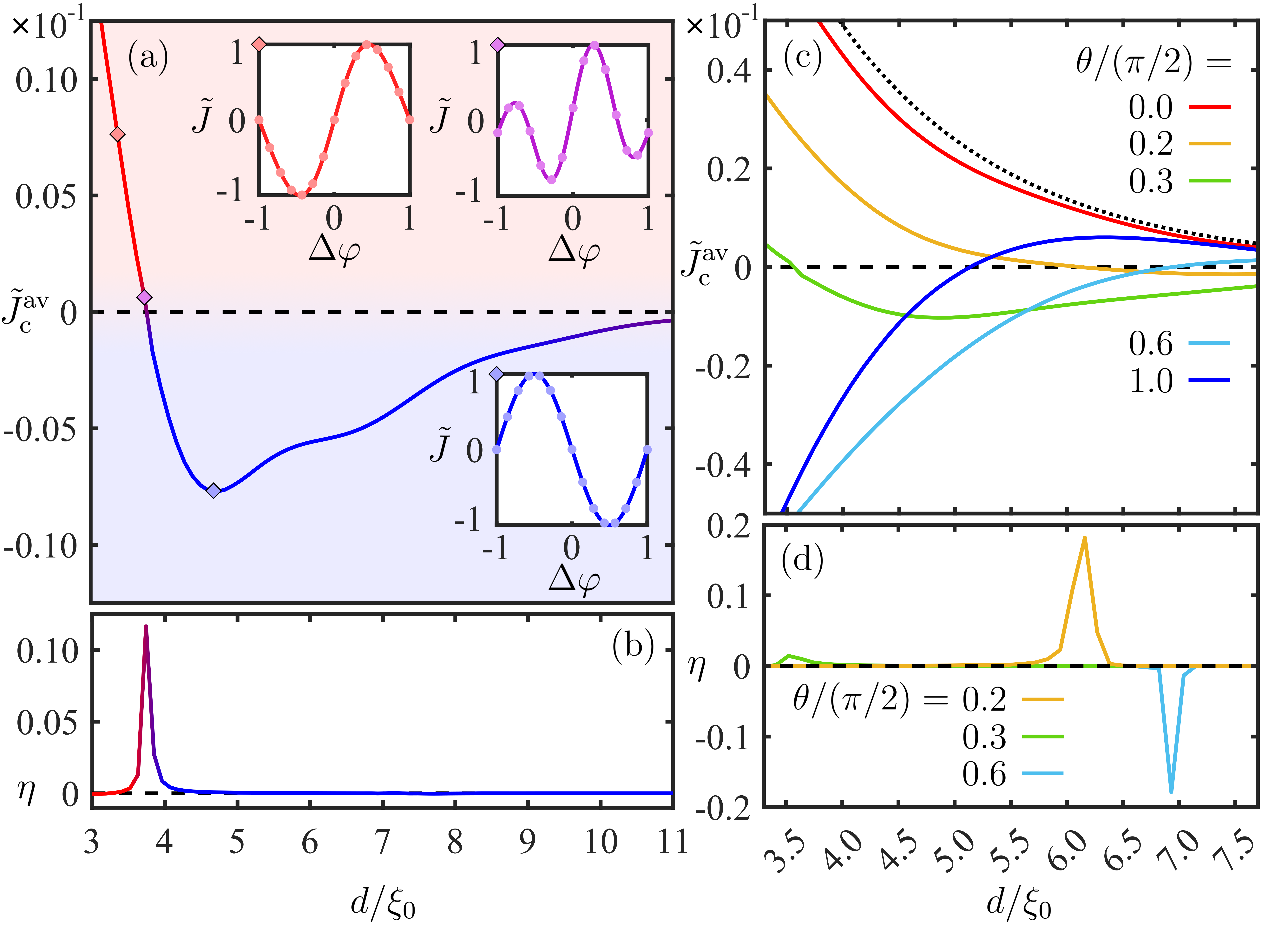}
    \caption{(a) When increasing the distance $d$ between the two superconductors with respect to their coherence length $\xi_0$, the average critical current [$\tilde{J}_{\text{c}}^{\text{av}}=J_{\text{c}}^{\text{av}}(d)/J_{\text{c}}^{\text{av}}(d\to0)$] of the conical magnet Josephson junction undergoes transitions between the $0$ state ($\tilde{J}_{\text{c}}^{\text{av}}>0$, upper left inset) and the $\pi$ state ($\tilde{J}_{\text{c}}^{\text{av}}<0$, lower inset). Close to the transition, the amplitude $J_2$ of the second harmonic in Eq.~\eqref{eq:Jphi} is comparable to $J_1$ ($|\tilde{J}_{\text{c}}^{\text{av}}|\ll1$, upper right inset), resulting in (b) a finite diode efficiency $\eta$. In the insets of panel~(a), the curves representing $\tilde{J}=J(\Delta\varphi)/|J_{\text{c},+}|$ are fits to the data (dots) using Eq.~\eqref{eq:Jphi}. (c) While a Josephson junction with a helical magnet ($\theta=0$) is in the $0$ state with $\tilde{J}_{\text{c}}^{\text{av}}$ only slightly suppressed compared to a normal metal with $h=0$ (black dotted curve), the Josephson junction undergoes $0-\pi$ transitions when increasing the tilt angle $\theta$ due to the increase in the net spin-splitting field. (d) A finite diode efficiency appears close to the $0-\pi$ transitions when $0<\theta<\pi/2$. We consider a system of size $(N_x,N_y) =(20+d/a,200)$, $d/a\in\{30,70\}$ with a clockwise spin rotation with period $\lambda_h=2.8\xi_0$, $\xi_0 = 9.1a$, $\Delta_0=0.07t$, $h=\Delta_0$, and $T=0.5T_{\text{c}}$. In panel~(a)-(b) $E_{\text{F}}=0.9t$ ($f=0.30$) and $\theta = 0.6\pi/2$. In panel~(c)-(d) $E_{\text{F}}=1.3t$ ($f=0.35$).}
    \label{fig:02}
\end{figure}

\prlsection{Diode efficiency}
When the magnet has a net spin-splitting field, the average critical current $J_{\text{c}}^{\text{av}}=(J_{\text{c},+}+J_{\text{c},-})/2$ can take positive or negative values, which if the critical current is sufficiently large,  correspond to a $0$ or $\pi$ phase, respectively.
Close to the $0-\pi$ transition, the amplitude of the second harmonic $J_2$ grows larger compared to $J_1$ [Eq.~\eqref{eq:Jphi}], resulting in nonreciprocity in the critical currents $J_{\text{c},+}\neq J_{\text{c},-}$ between positive and negative $\Delta\varphi$, and thus a finite diode efficiency, see Fig.~\ref{fig:02}(a)-(b). While the $0-\pi$ transition can be approached by designing junctions with an appropriate distance between the superconductors, further tuning is achieved by increasing the out-of-plane component of the spin-splitting field $h\sin(\theta)$ via an increase in the tilt angle $\theta$ [Fig.~\ref{fig:02}(c)-(d)]. This leads to a more rapid oscillation of the average critical current as a function of the distance $d$ between the superconductors \cite{Volkov_PRB_2006}. An overall increase in the local spin splitting field $h$ has a similar effect (see SM). The diode efficiency is odd under inversion of $h\sin(\theta)$ and the rotation direction ($\lambda_h\to-\lambda_h$).

\begin{figure}[tb]
    \centering
    \includegraphics[width=\columnwidth]{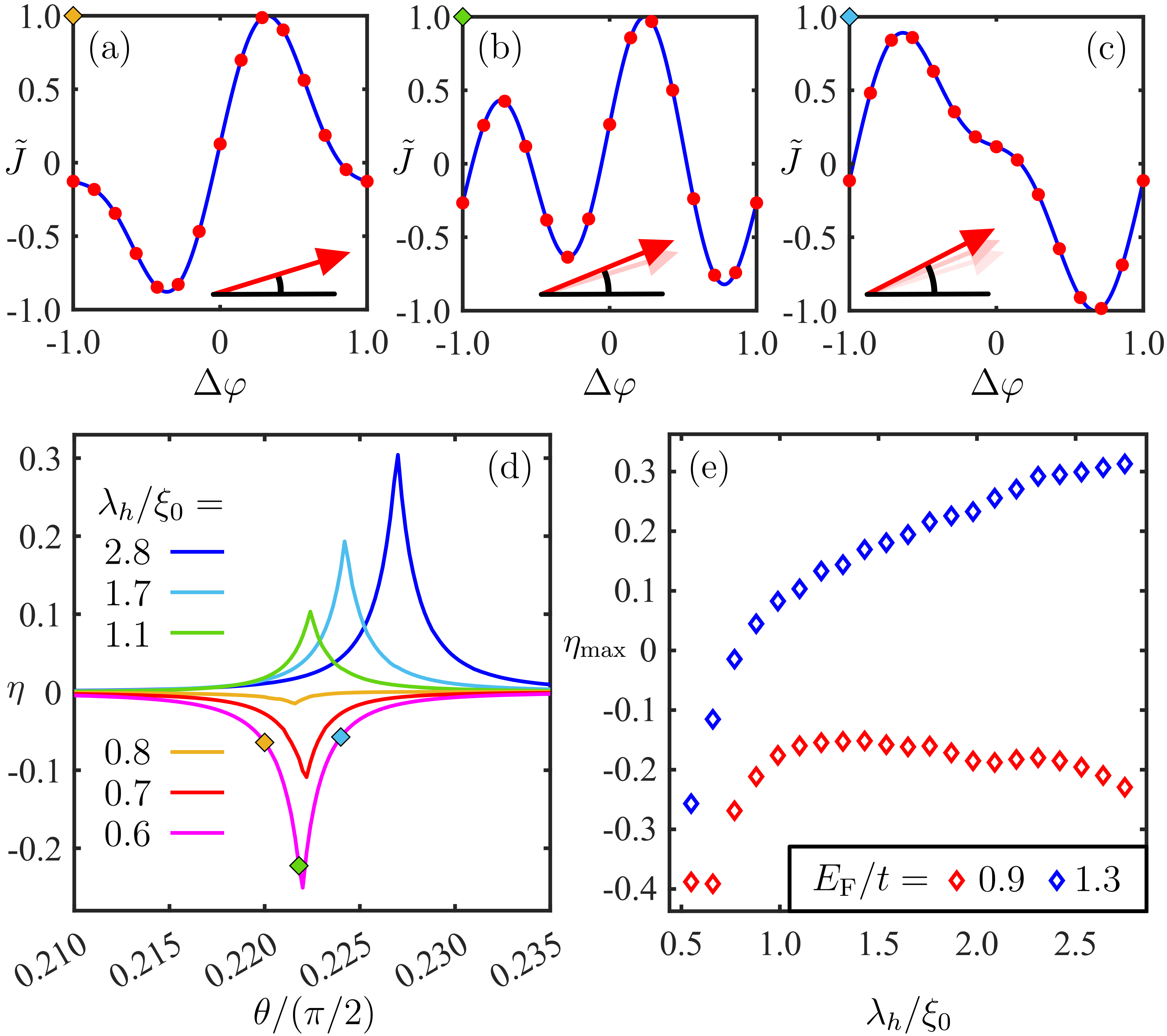}
    \caption{(a)-(c) When the tilt angle increases, as schematically illustrated by the red arrows (not to scale), the Josephson junction undergoes a $0-\pi$ transition, close to which the current-phase relation $\tilde{J}=J(\Delta\varphi)/|J_{\text{c,}+}|$ reveals nonreciprocal critical currents $J_{\text{c,}+}\neq J_{\text{c,}-}$. Here, $\lambda_h=0.6\xi_0$ as marked by the color coded squares in panel~(d). The blue curves are fits to the data (red dots) using Eq.~\eqref{eq:Jphi}. (d) The diode efficiency $\eta$ is plotted as a function of the tilt angle $\theta$ for various values of the rotation period $\lambda_h$ with respect to the coherence length $\xi_0$ for $E_{\text{F}}=1.3t$. (e) The peak value of the diode efficiency $\eta_{\text{max}}=\text{sgn}(\eta)\text{max}(|\eta|)$ is plotted as a function of $\lambda_h/\xi_0$. For all panels, we consider a system of size $(N_x,N_y)=(65,200)$ with a clockwise spin rotation, $\Delta_0=0.07t$, $\xi_0 = 9.1a$, $T=0.5T_{\text{c}}$, and $h=\Delta_0$. The Fermi energies $E_{\text{F}}=0.9t$ and $1.3t$ correspond to $f = 0.30$ and $0.35$, respectively.}
    \label{fig:03}
\end{figure}

In Fig.~\ref{fig:03}, we further explore the diode effect close to the field-tunable $0-\pi$ transition.
A large diode efficiency can be achieved when the rotation period $\lambda_h$ is of the same order of magnitude as the coherence length of the superconductor. 
The rotation period is $\lambda_h=48$~nm for Cr$_{1/3}$NbS$_2$ \cite{Togawa_PRL_2012} with a 9$^{\circ}$ rotation between nearest neighbor sites \cite{Cao_MaterTodayAdv_2020}. A coherence length $\xi_0\sim\mathcal{O}(10~\text{nm})$ of the same order of magnitude is realizable, \textit{e.g.}, as recently studied in bilayers consisting of Cr$_{1/3}$NbS$_2$ and superconducting NbS$_2$ \cite{Spuri_PhysRevResearch_2024}. This experimental work found signatures of long-range spin-triplet Cooper pairs, suggesting that transport through Josephson junctions much longer than the coherence length of the parent spin-singlet superconductor [Fig.~\ref{fig:02}] is feasible. Similar long-range transport was recently observed also in antiferromagnetic Josephson junctions owing to a noncollinear spin structure \cite{Jeon_NatMater_2021,Jeon_NatNanotech_2023}.
The nonmonotonic behavior in Fig.~\ref{fig:03}(e) relates to the sensitivity to changes in other parameters than $\lambda_h$ due to a small oscillatory contribution to the critical current through the conical magnet \cite{Halasz_PRB_2009}, see Fig.~\ref{fig:02}(a). This oscillatory term shifts the $0-\pi$ transition towards higher or lower tilt angles $\theta$ when $\lambda_h$ increases, corresponding to a change in the net spin-splitting field $h\sin(\theta)$.

\prlsection{Concluding remarks}
We have thus shown that the conical spin structure found in, \textit{e.g.}, Ho \cite{Sosnin_PRL_2006} and tilted Cr$_{1/3}$NbS$_2$ \cite{Cao_MaterTodayAdv_2020} can produce a Josephson diode with considerable diode efficiencies close to the $0-\pi$ transition. The inversion symmetry is broken by the helical spin rotation that gives rise to quasi-1D Rashba-like band splitting inversely proportional to the rotation period. Time reversal symmetry is broken by the tilt that creates a noncoplanar spin texture. A Josephson diode can thus be realized using a single magnetic material, without relying on spin-orbit coupling. While external magnetic fields are not required, they can provide a useful knob for tuning the Josephson diode effect.

%----------------------------------------------------
% Acknowledgements:

\begin{acknowledgements}

We thank Yugo Onishi, Margarita Davydova, Jagadeesh Moodera and Yasen Hou  for helpful discussions on superconducting diodes. We thank Nadya Mason and Suyang Xu for stimulating discussions on conical magnets. This work was supported by Simons Investigator Award from Simons Foundation.  

\end{acknowledgements}

%----------------------------------------------------

% Bibliography:

%\bibliography{references.bib}


%----------------------------------------------------

%SM attaced:

\onecolumngrid
\newpage
\begin{center}
\noindent
{\large\textbf{Supplemental material to: Nonreciprocal Josephson current through a conical magnet}}\\
\vspace{1em}Lina Johnsen Kamra and Liang Fu\\
\vspace{0.5em}{\small$^1$\textit{Department of Physics, Massachusetts Institute of Technology, Cambridge, MA 02139, USA}\\
(Dated: \today)}
\end{center}
\vspace{2em}

\noindent
We here provide i) further details about the Bogoliubov--de Gennes equations that we solved numerically, ii) results for the average critical current at different values of the local spin-splitting field $h$, and iii) results for the diode efficiency for different Fermi energies.
\vspace{2.5em}

\twocolumngrid
\setcounter{equation}{0}
\renewcommand{\theequation}{S.\arabic{equation}}
\setcounter{figure}{0}
\renewcommand{\thefigure}{S.\arabic{figure}}

\section{Bogoliubov--de Gennes equations}

\begin{figure}[b]
    \centering
    \includegraphics[width=0.55\columnwidth]{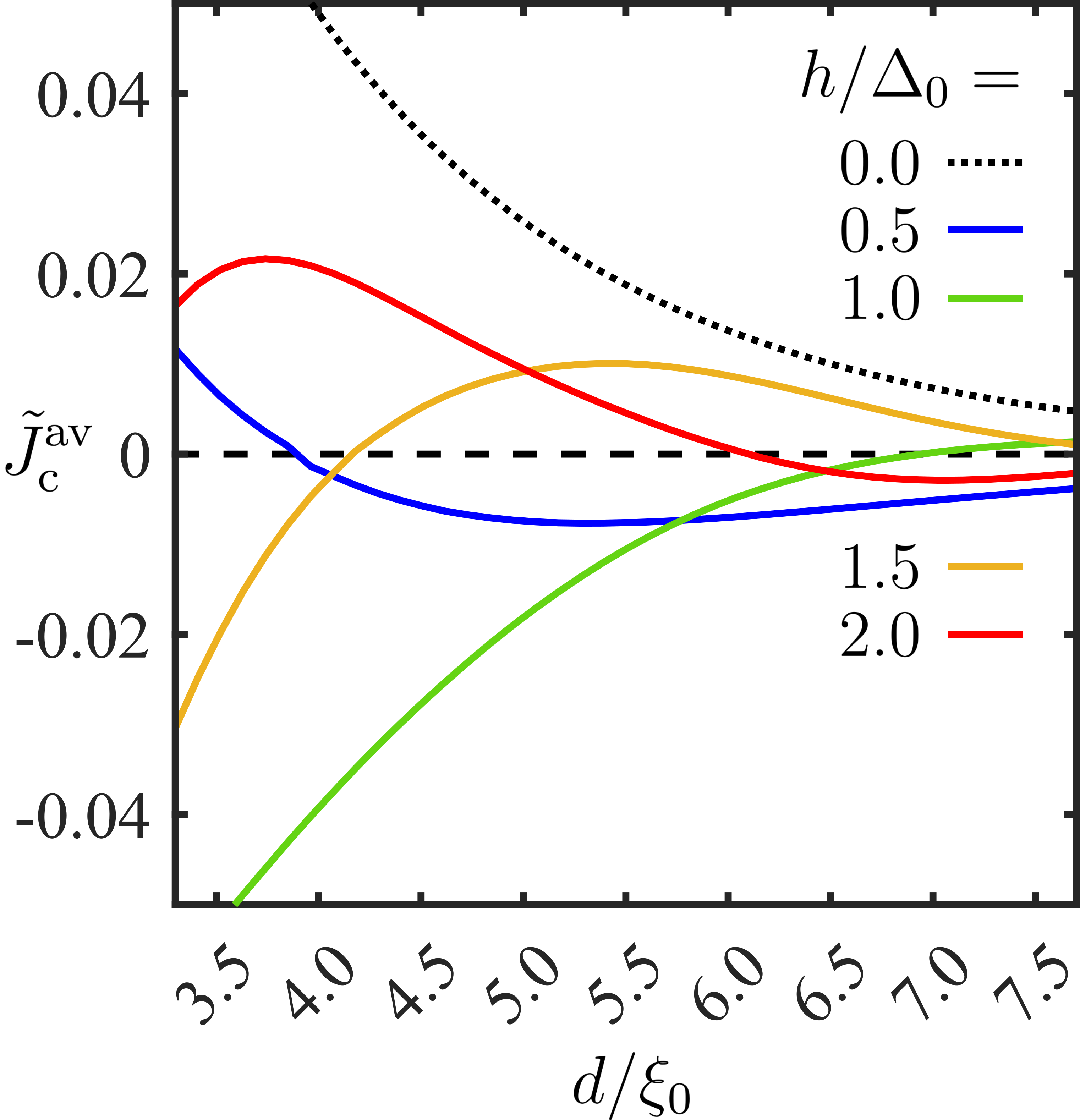}
    \caption{The conical magnet Josephson junction undergoes $0-\pi$ transitions when increasing the spin-splitting field $h$ and the distance $d$ between the two superconductors with respect to the coherence length $\xi_0$. We consider a system of size $(N_x,N_y) =(20+d/a,200)$, $d/a\in\{30,70\}$ with a clockwise spin rotation with period $\lambda_h=2.8\xi_0$, tilt angle $\theta = 0.6\pi/2$, $E_{\text{F}}=1.3t$ ($f=0.35$), $\Delta_0=0.07t$, $\xi_0 = 9.1a$, $h=\Delta_0$, and $T=0.5T_{\text{c}}$.}
    \label{fig:SM01}
\end{figure}

We here provide details on the numerical method \cite{Zhu_Book_2016,Chourasia_PRB_2023} used to obtain Figs.~2 and~3 in the main text.
Starting from the Hamiltonian in Eq.~(5) in the main text, we assume periodic boundary conditions in the $y$ direction and apply the Fourier transform
\begin{align}
    c_{\ve{i},\sigma}=\frac{1}{\sqrt{N_y}}\sum_{k_y}c_{i_x,k_y,\sigma}e^{ik_y ai_y}.
\end{align}
By using the relation
\begin{align}
    \frac{1}{N_y}\sum_{i_y}e^{i(k_y-k_y^{'})ai_y}=\delta_{k_y,k_y^{'}},
\end{align}
and defining a basis 
\begin{align}
    \psi_{i_x,k_y}=[c_{i_x,k_y,\uparrow}\:\:c_{i_x,k_y,\downarrow}\:\:c_{i_x,-k_y,\uparrow}^{\dagger}\:\:c_{i_x,-k_y,\downarrow}^{\dagger}]^T,
\end{align}
we can write the Hamiltonian in the form
\begin{align}
    H = \frac{1}{2}\sum_{i_x,j_x,k_y}\psi_{i_x,k_y}^{\dagger}H_{i_x,j_x,k_y}\psi_{j_x,k_y}.
\end{align}
We have defined a Hamiltonian matrix
\begin{align}
    H_{i_x,j_x,k_y} = (&\delta_{i_x+1,j_x}+\delta_{i_x-1,j_x})[-t\:\text{diag}(\sigma_0,-\sigma_0)]\notag\\
    +&\delta_{i_x,j_x}\{[-2t\cos(k_y)-\mu]\:\text{diag}(\sigma_0,-\sigma_0)\notag\\
    +&\ve{h}_{i_x}\cdot\text{diag}(\ve{\sigma},-\ve{\sigma}^*)\notag\\
    +&\text{antidiag}(\Delta_{i_x},-\Delta_{i_x},-\Delta_{i_x}^*,\Delta_{i_x}^*)\}.
\end{align}
To obtain the corresponding eigenenergies $E_{n,k_y}$ and eigenvectors
\begin{align}
    \phi_{n,i_x,k_y}=[u_{n,i_x,k_y,\uparrow}\:\:u_{n,i_x,k_y,\downarrow}\:\:v_{i_x,k_y,\uparrow}\:\:v_{i_x,k_y,\downarrow}]^T,
\end{align}
we numerically solve the Bogoliubov--de Gennes equations \cite{Zhu_Book_2016}
\begin{align}
    \sum_{j_x}H_{i_x,j_x,k_y}\phi_{n,j_x,k_y}=E_{n,k_y}\phi_{n,i_x,k_y}.
\end{align}
By realizing that there is a second equivalent solution
\begin{align}
    -E_{n,-k_y};\:[v^*_{n,i_x,-k_y,\uparrow}\:\:v^*_{n,i_x,-k_y,\downarrow}\:\:u^*_{i_x,-k_y,\uparrow}\:\:u^*_{i_x,-k_y,\downarrow}]^T
\end{align}
we can write the Hamiltonian as 
\begin{align}
    H = \sum_{n,k_y}^{'}E_{n,k_y}\gamma_{n,k_y}^{\dagger}\gamma_{n,k_y},
\end{align}
where $\sum_{n,k_y}^{'}$ is the sum over positive eigenenergies $E_{n,k_y}>0$ only, and $\gamma_{n,k_y}$ are the new fermion operators. We have disregarded constant terms.
Physical observables can be evaluated by expressing the old fermion operators in terms of the new ones using the relation
\begin{align}
    c_{i_x,k_y,\sigma} = \sum_n^{'}\big[u_{n,i_x,k_y,\sigma}\gamma_{n,k_y}+v^*_{n,i_x,-k_y,\sigma}\gamma_{n,-k_y}^{\dagger}],
\end{align}
and by evaluating expectation values of the new operators as
\begin{align}
    \langle &\gamma_{n,k_y}^{\dagger}\gamma_{n^{'},k_y^{'}}\rangle = \delta_{n,n^{'}}\delta_{k_y,k_y^{'}} f_{\text{FD}}(E_{n,k_y}),\\
    \langle &\gamma_{n,k_y}^{\dagger}\gamma^{\dagger}_{n^{'},k_y^{'}}\rangle = \langle \gamma_{n,k_y}\gamma_{n^{'},k_y^{'}}\rangle = 0,
\end{align}
where $f_{\text{FD}}(E_{n,k_y})$ is the Fermi-Dirac distribution.
The filling fraction given in Eq.~(7) in the main text thus takes the form
\begin{align}
    f &=\frac{1}{N_x N_y}\sum_{i_x,\sigma}\sum_{n,k_y}^{'}\big\{\left|u_{n,i_x,k_y,\sigma}\right|^2f_{\text{FD}}(E_{n,k_y})\notag\\
    &\hspace{7em}+\left|v_{n,i_x,k_y,\sigma}\right|^2\left[1-f_{\text{FD}}(E_{n,k_y})\right]\big\},
\end{align}
and the $x$ oriented local bond current from site $i_x$ to its nearest neighbor $i_x+1$ defined in Eq.~(8) in the main text is given by
\begin{align}
    &J^x_{i_x+1,i_x}=\frac{it}{N_y}\sum_{\sigma}\sum_{n,k_y}^{'}\big\{u_{n,i_x+1,k_y,\sigma}^* u_{n,i_x,k_y,\sigma}f_{\text{FD}}(E_{n,k_y})\notag\\
    &+v_{n,i_x+1,k_y,\sigma}v_{n,i_x,k_y,\sigma}^*[1-f_{\text{FD}}(E_{n,k_y})]-\text{c.c}\big\}.
\end{align}

\begin{figure}[tb]
    \centering
    \includegraphics[width=\columnwidth]{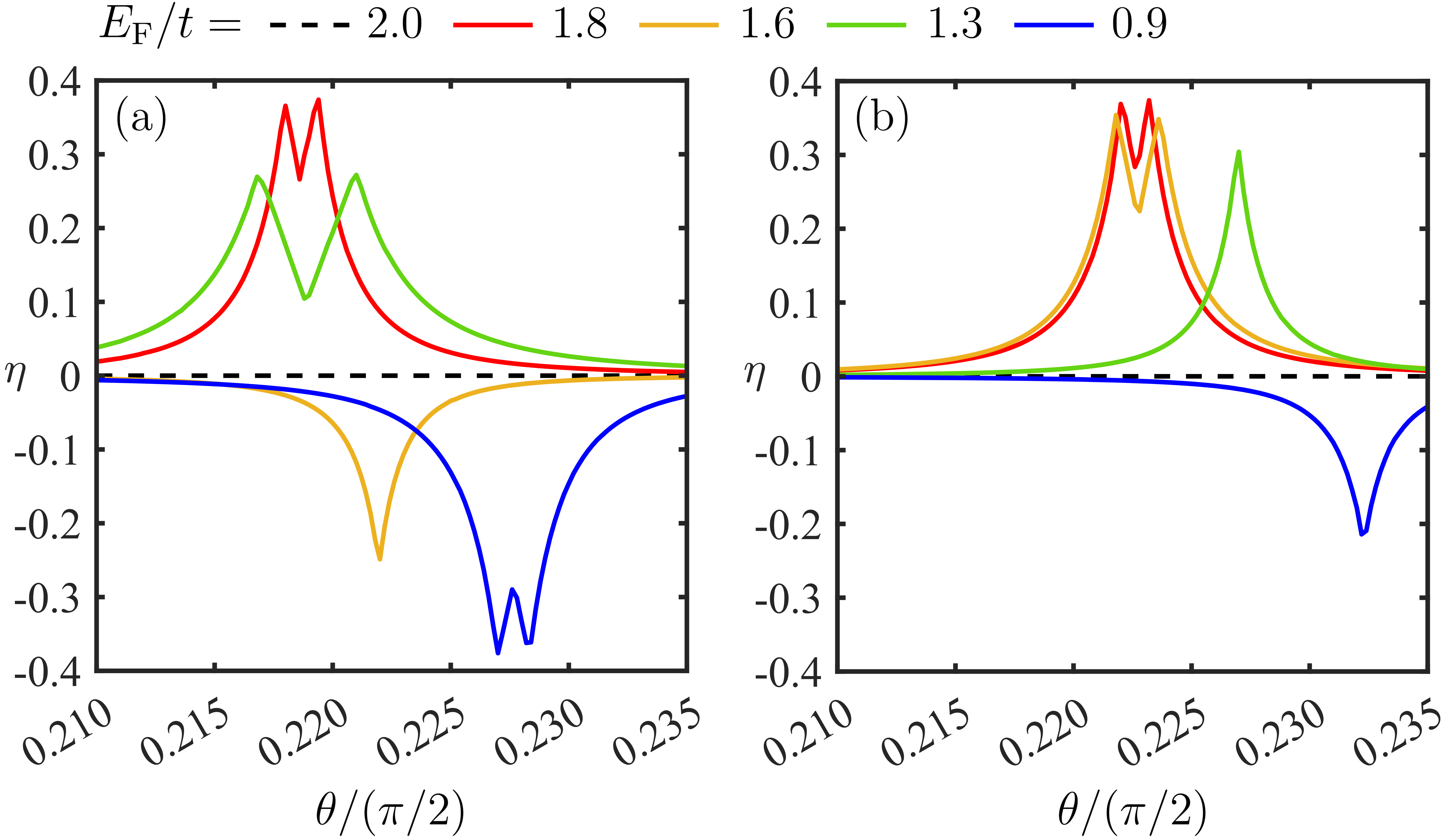}
    \caption{The diode efficiency $\eta$ is plotted as a function of the tilt angle $\theta$ for various values of the Fermi energy $E_{\text{F}}$ for (a) $\lambda_h=0.6\xi_0$ and (b) $\lambda_h=2.8\xi_0$. The Fermi energies $E_{\text{F}}=(2.0t,1.8t,1.6t,1.3t,0.9t)$ correspond to $f = (0.50,0.45,0.40,0.35,0.30)$, respectively. We consider a system of size $(N_x,N_y)=(65,200)$ with a clockwise spin rotation, $\Delta_0=0.07t$, $\xi_0 = 9.1a$, $T=0.5T_{\text{c}}$, and $h=\Delta_0$.}
    \label{fig:SM02}
\end{figure}

\section{Additional results}

In the main text, we showed that the magnetic Josephson junction with a conical spin structure undergoes $0-\pi$ transitions when increasing the out-of-plane component of the spin-splitting field $h\sin(\theta)$.
In Fig.~\ref{fig:SM01}, we show that the Josephson junction also undergoes $0$-$\pi$ transitions when increasing the magnitude of the local spin-splitting field $h$ \cite{Buzdin_JETPLett_1991,Ryazanov_PRL_2001,Birge_APLMater_2024}. The frequency of the oscillations of the average critical current as a function of the distance $d$ between the two superconductors increases with increasing $h$.

The diode efficiency is zero at half-filling ($f=1/2$) when the dispersion is neither electron-like nor hole-like. When the dispersion obtains a finite curvature away from half filling, a finite diode efficiency appears close to the $0-\pi$ transition as shown in Fig.~\ref{fig:SM02}. The diode efficiency is odd in the deviation from half filling $f-1/2$, and thus takes opposite signs for electron-like and hole-like bands. In the main text, we have considered a filling well below half filling in order to approach the electron-like quadratic dispersion in Fig.~\ref{fig:01}(d)-(e). 

%----------------------------------------------------

% Acknowledgements:

%\begin{acknowledgements}

%\end{acknowledgements}

%----------------------------------------------------

% Bibliography:

%\bibliography{references.bib}

\end{document}